\documentclass[12pt,preprint]{aastex}

\shorttitle{Temperature Variations in NGC 7009}
\shortauthors{Rubin et al.}

\begin{document}

\title{Temperature Variations from HST Imagery and Spectroscopy of 
NGC~7009\altaffilmark{1}}


\author{R.~H.~Rubin\altaffilmark{2,3}, 
N.~J.\ Bhatt\altaffilmark{2},  
R.~J.~Dufour\altaffilmark{4}, 
B.~A.\ Buckalew\altaffilmark{4},  
M.~J.\ Barlow\altaffilmark{5},  
X.-W.\ Liu\altaffilmark{5},  
P.~J.\ Storey\altaffilmark{5},  
B.\ Balick\altaffilmark{6},  
G.~J.~Ferland\altaffilmark{7}, 
J.~P.\ Harrington\altaffilmark{8}, 
and
P.~G.~Martin\altaffilmark{9}}
\altaffiltext{1}{Based on observations made with the NASA/ESA {\it Hubble Space
Telescope}, obtained at ST~ScI, which is 
operated by AURA, Inc., under NASA contract NAS5-26555}  
\altaffiltext{2}{NASA/Ames Research Center, Moffett Field, CA 94035-1000}
\altaffiltext{3}{Orion Enterprises}
\altaffiltext{4}{Rice University, Physics \& Astronomy, MS 61, 
Houston, TX 77005-1892}
\altaffiltext{5}{University College London, Physics \& Astronomy, Gower 
Street, London, UK WC1E~6B}
\altaffiltext{6}{University of Washington, Astronomy, Seattle, WA 98195-1580}
\altaffiltext{7}{University of Kentucky, Physics \& Astronomy, 
Lexington, KY 40506-0055}
\altaffiltext{8}{University of Maryland, Astronomy, College Park, MD 20742-2421}
\altaffiltext{9}{Canadian Institute for Theoretical Astrophysics, University 
of Toronto, Toronto, ON, Canada M5S~3H8}



\begin{abstract}
We present new HST/WFPC2 imagery and STIS long-slit spectroscopy
of the planetary nebula NGC~7009.
The primary goal was to obtain high spatial resolution
of the intrinsic line ratio [\ion{O}{3}] 
4364/5008 and thereby evaluate the
electron temperature ($T_e$) and the fractional mean-square $T_e$ variation 
($t_A^2$) {\it across the nebula}.
The WFPC2 $T_e$ map is rather uniform;
almost all values are between 9000~-- 11,000~K, with the higher $T_e$s
closely coinciding with the inner He$^{++}$-zone.
The results indicate very small values~-- $\la$0.01~--  
for $t_A^2$ throughout.
Our STIS data allow an even more direct determination of $T_e$ and 
$t_A^2$, albeit for a much smaller area than with WFPC2.
We present results from binning the data along the slit into tiles 
that are 0.5$''$ square (matching the slit width).
The average [\ion{O}{3}] temperature using 45 tiles 
(excluding the central star and
STIS fiducial bars) is 10,139~K; $t_A^2$ is 0.0035.
The measurements of $T_e$ reported here are an average
along each line of sight.
Therefore, despite finding remarkably low $t_A^2$, we cannot completely
rule out temperature fluctuations along the line of sight as the cause 
of the large abundance discrepancy between heavy element abundances 
inferred from collisionally excited emission lines compared to those 
derived from recombination lines.

\end{abstract}

\vskip-0.2truein


\keywords{ISM: planetary nebulae --- ISM: abundances --- ISM: atoms ---
ISM: individual (NGC~7009)}

%% \keywords{clusters: globular, peanut---bosons: bozos}


\section{INTRODUCTION}

     Most observational tests of the chemical evolution of the universe
rest on emission line objects; these define the endpoints of stellar
evolution and probe the current state of the interstellar medium.
Gaseous nebulae are laboratories for understanding physical 
processes in all emission-line sources, and probes for stellar, 
galactic, and primordial nucleosynthesis. 

	There is a fundamental issue that continues to be problematic~--
the discrepancy between heavy element
abundances inferred from emission lines that are collisionally excited 
compared with those 
due to recombination/cascading, the so-called ``recombination lines''.
Studies of planetary nebulae (PNs)
contrasting recombination and collisional abundances 
(Liu et~al.\ 1995, Kwitter \& Henry 1998)
often find differences exceeding a factor of two. 
In an extensive study of NGC~6153, Liu et~al.\ (2000) found
that C$^{++}$/H$^+$, 
N$^{++}$/H$^+$, 
O$^{++}$/H$^+$, and
Ne$^{++}$/H$^+$ ratios derived from optical recombination lines
are all a factor of $\sim$10 higher than the corresponding values
deduced from collisionally-excited lines.
Abundances determined from these two methods 
disagree by a factor larger than 
the spread of abundances used to determine such fundamental
quantities as Galactic abundance gradients (e.g., Shaver et~al.\ 1983; Simpson
et~al.\ 1995; Henry \& Worthey 1999).

	Most of the efforts to explain the abundance 
puzzle between collisional and recombination values have attempted
to do so by examining electron temperature ($T_e$) variations in the plasma.
This is often done, using the formalism of Peimbert (1967), in terms
of the [fractional] mean-square variation ($t^2$) of $T_e$.
The inferred metallicity obtained by using the usual (optical/UV) 
forbidden lines is very sensitive to $T_e$ (exponential) and $t^2$.
On the other hand, recombination lines 
are rather insensitive to $T_e$ and $t^2$.
Agreement  close to the higher recombination value
can be forced in the derived abundance 
by attributing the difference to (solving for) $t^2$.
Let us consider the case of the PN NGC~7009,
also known as the ``Saturn Nebula"; its image produced
by Balick et~al.\ (1998) from WFPC2 data has graced the cover
of coffee-table books (Petersen \& Brandt 1998).
NGC~7009 has stood near the centre of the abundances controversy.
It has long been known that its recombination lines are  
very strong, particularly those of \ion{O}{2}. 
Liu et~al.\ (1995) found 
that the recombination C, N, and O abundances are a factor of $\sim$5
larger than the corresponding collisional abundances.
This was recently found to be the case also for neon
(Luo, Liu \& Barlow 2001).
One possible reconciliation of these two drastically different abundances,
which Liu et~al.\ (1995) discuss, is to invoke large $t^2$.
For oxygen in NGC~7009,
by invoking $t^2$~$\sim$ 0.1, 
agreement can be forced in the 
abundance derived from optical lines
close to the higher recombination value~--
a value more than 2.5 times larger than the solar
O/H of 7.41$\times$10$^{-4}$ (Grevesse \& Sauval 1998).
Such a large $t^2$ is not at all predicted by current theory/models
(e.g., Kingdon \& Ferland 1998).

	The current unsettled situation
has led to efforts to broaden the study to include other 
variables besides $T_e$ to 
analyse 
the effects upon
abundance determinations.
One promising avenue is to examine this considering density variations,
abundance variations, and $T_e$ variations in combination.
Liu et~al.\ (2000) took this approach in their investigation of
the PN NGC~6153 with a two-phase empirical model.
Recall this object has even more disparate abundance differences 
(factor of $\sim$10) than does NGC~7009.
P\'equignot et~al.\ (2002) continued the study using
photoionization models including two components with different 
heavy element abundances.
Basically, one phase is small, dense, ionized clumps that are highly 
enhanced in heavy elements and deficient in hydrogen.
This phase would have $T_e$ $\sim$10$^3$~K and contribute
almost all the heavy element recombination line emission.
These clumps are embedded in the second phase which has the 
more usual nebular properties/composition,
with $T_e$ $\sim$10$^4$~K,
and which is responsible for almost all the
emission in the optical/UV collisionally excited lines.
The latter phase comprises the great bulk of the nebular mass.
There is ongoing work by this same group to try to explain
other PNs of this kind, including
the Galactic bulge PN M~1-42, which has 
a factor of $\sim$20 abundance dichotomy 
(Liu et~al.\ 2001) and Hf~2-2, which has the most extreme 
abundance difference to date, a factor of $\sim$80
(Liu 2002).

	In this paper, we focus solely on $T_e$ variations with the purpose 
to test whether the observational data support
the possibility that $t^2$ in NGC~7009 may be as large as $\sim$0.1.
In section 2, we present the {\it HST} observations
and data reduction procedures.
Section 3 describes our techniques for obtaining emission
line fluxes from the WFPC2 data and how we use cospatial
STIS data to help this process.
In section 4, there is a discussion and analysis of extinction.
We determine the electron temperature distributions in section 5.
In section 6, we analyse the $T_e$ distributions 
in terms of average temperatures and fractional mean-square 
temperature variations in the plane of the sky.
Section 7 provides a discussion and conclusions.

\section{{\it HST} OBSERVATIONS AND DATA REDUCTION}

	The observations of NGC~7009 described here were taken as part of
our HST Cycle 8 program GO-8114.
For both the WFPC2 and STIS visits, we set our positions with respect to
the position of the central star (CS):
$\alpha$, $\delta$ = $21^{\rm h}04^{\rm m}10\fs81$, 
--11$^{\rm o}$21\arcmin47\farcs91)
(equinox J2000).

\subsection{WFPC2 Observations}

	We observed with WFPC2 on 2000 April 6--7 (UT) as the first
visit of the GO-8114 program. 
Observations (with corresponding line of interest)
of NGC~7009
were made in line filters F437N ([\ion{O}{3}] 4364~\AA),
F487N (H$\beta$ 4863~\AA),  F502N ([\ion{O}{3}] 5008~\AA),
and F656N (H$\alpha$ 6565~\AA),  
plus continuum filter F547M.
All wavelengths in this paper are vacuum rest wavelengths.
The images were placed in the WF3 chip (with the CS at WF3-FIX).
The pixel size is 0.0996$''$.
In order to clean cosmic rays, two images were taken in each filter
except for three images in F437N.
Total integration times were respectively: 2200, 520, 320, 400, and
240~s. 

The data sets we processed (also for STIS)
were those obtained after sufficient time elapsed 
from the observation dates in order that ``best reference files" would
be stable/finalized.  We requested On-the-Fly Calibration for science files
and Best Reference Files.
We then co-added and cosmic ray cleaned images using standard packages
in IRAF.\footnote{IRAF is distributed by NOAO, which is operated by AURA,
under cooperative agreement with NSF.}
Flux calibration for each filter was done by basically following the procedures
used for the WFPC2 Exposure Time Calculator.  These account for 
the integration time, analog-to-digital converter gain of 7 
(used for F437N and  F487N) or 14 electrons/DN (used for F502N, F547M, 
and F656N),  and the total system throughput.

	The calculations for the contributions  to the narrow-band
filter observations  of the continuum and other lines will be 
discussed later (\S 3) after the STIS observations section.

\subsection{STIS Observations}

	On 2000 August 9 (UT), STIS long-slit spectra of NGC~7009 were taken.  
These data comprise 3 orbits and were taken at position angle
(PA)~= 86.72$^{\rm o}$.
Our original plan was to align the slit with the major axis of
NGC~7009, which is at PA~= 78.5$^{\rm o}$.
However,
we were advised by STScI that 
there were no guide star pairs for our planned visits.  
We settled on the revised PA, 
a change of only 8.22$^{\rm o}$,
as this was the closest to the major axis
that allowed for a pair of guide stars.
We used a slit width of 0.5$''$ and a slit length of 52$''$ for these 
CCD observations.

	Because the central star (CS) in NGC~7009 is reasonably bright
(V~= 12.78, B~= 12.66), it 
would have caused 
problems (bleeding, saturation, etc.), 
especially on long exposures, if it were in our science slit.  
On the other hand, the CS provides an excellent fiducial point in the nebula, 
where we may assess the positional registration of emission lines in 
our various exposures with STIS.
Thus, we observed with the CS off the upper edge of our 52$''$$\times$0.5$''$
slit by 0.1$''$.
In other words, our slit is defined with the CS moved 0.35$''$ (0.25 + 0.1)
in the dispersion direction ($\sim$NNW).  
Furthermore the slit was not centred on the CS in the spatial direction.
Because the structure on the west-side of the nebula was of more
interest to us, we displaced the slit centre 3.5$''$ from the CS 
in the spatial direction ($\sim$W along PA~= 86.72$^{\rm o}$).
Spectra were taken with gratings G430M (with 
wavelength settings: 3680, 4194, 4451, 4706, 4961) 
and G750M (settings: 5734, 6581).  
Each exposure was done in accumulation mode and at least
two spectra were taken at each setting in order to cosmic-ray (CR) clean.
After retrieving the data sets, we then co-added and cosmic ray 
cleaned images using standard packages in IRAF.
Calibrations to produce 2-dimensional (2D) rectified images 
were then carried out.
 From these, we singled out specific emission lines for further investigation.

	Data for the [\ion{O}{3}] 4364~\AA\ line as well as H$\gamma$ 4342~\AA\
were contained in the G430M/4451 grating setting (150s);
for the [\ion{O}{3}] 5008, 4960~\AA\ lines and H$\beta$ 4863~\AA\
lines, G430M/4961 was used (80s);
for H$\alpha$ 6565~\AA\ and the
[\ion{N}{2}] 6550, 6585~\AA\ lines, G750M/6581 was used (90s).
For G430M, the dispersion is 0.28~\AA\ per pixel 
for a point source and the plate scale 0.05 arcsec/pixel
(Leitherer et~al.\ 2001, Chapter~13).
For a uniformly filled slit
with width 0.5$''$, a degradation in
resolving power by a factor of 10 is expected
to a spectral resolution of 2.8~\AA.  
For G750M, the dispersion is 0.56~\AA\ per pixel 
for a point source and the plate scale 0.05 arcsec/pixel.
For a uniformly filled slit
with width 0.5$''$, the spectral resolution would be 5.6~\AA.  
	For the analysis presented here, we were interested mainly in
the distribution of line flux along the slit spatial direction.
This was accomplished with the IRAF routine {\it blkavg} in conjunction
with specialized software tools that we developed ourselves.

	Even after applying the standard CR rejection 
there still remain many bad pixels due to CRs and/or hot pixels.
There is considerable danger that including these can corrupt 
the flux values we seek.  The program developed to eliminate these
remaining bad pixels is called PIXHUNTER, which is described briefly 
in Appendix~A. 
Once the columns containing the line have been cleaned for 
deviant pixels, we are ready to subtract an 
equivalent spectral range of continuum.
We do this by using IRAF functions, including {\it blkavg},
to operate upon the appropriate sections of cleaned continuum. 
The 1-dimensional (1-D) distribution of line flux vs.\ spatial coordinate
for the various emission lines of interest is what we need
for our subsequent physical analysis.

	We note that there is excellent agreement with 
a cross check of the 1-D results of flux vs.\ spatial direction
by comparing with 1-D results of flux vs.\ wavelength for a corresponding
spatial sample.
The latter were measured with the {\it splot} package.
With this, the underlying continuum is fitted and the integrated
line flux determined with the $e$-option (area under the line profile), which
was preferable to fitting with a Gaussian profile.
Because of the spectral impurity introduced by the
relatively wide slit used, the line profiles have flatter tops and less 
extended bases (i.e., they are more ``trapezoidal'') 
than the Gaussian fits.  It is also apparent that
the Gaussian fit is overestimating the line flux.  

	Both the [\ion{O}{3}] 5008 and 4960~\AA\ lines were 
observed simultaneously with the\break
G430M/4961 grating setting.
We desired to accurately measure the ratio of these line fluxes
to compare with the most recent calculation for the theoretical intensity ratio
2.984 (Storey \& Zeippen 2000).
Because both transitions arise from the same upper level,
the intrinsic flux ratio depends only on the transition probabilities
(A-values) and wavelengths.
What we found was a surprising variation in the 
F(5008)/F(4960) ratio with position along the slit.
This amounts to a variation in the ratio of roughly 3.0$\pm$0.1.
Furthermore, it appears more-or-less 
periodic with an $\sim$3.5$''$ cycle.

	According to Ted Gull (private communication),
this is an instrumental effect and is a ratio of two fringe patterns. 
The source of the problem is a thin blocker filter that had to be matched 
with each grating and the best (and only) location that it could be placed was 
above the grating in a stable mounting. 
There is some discussion of fringing at the STIS web site
(http://www.stsci.edu/instruments/stis/performance/anomalies).
More specific information appears in chapter~7 of the STIS Instrument Handbook 
(Leitherer et~al.\ 2001)\break
under a section called ``Fringing due to the Order Sorter Filters".
See\break 
(www.stsci.edu/instruments/stis/documents/handbooks/cycle11/c07\_performance2.html\#326437).
This distinct fringing pattern also appears in similar STIS data of the
Orion Nebula obtained under program GO-7514 (PI RR).

	To attempt to do anything about fringing would probably require a
dedicated HST/STIS calibration program.  We do not treat this further here.
As will be seen, the results for the temperature analyses 
are in good accord between the two datasets using STIS and WFPC2,
which provides support that any fringing in the STIS data 
is not affecting our conclusions.
If there were fringing in the F(5008)/F(4364) ratio at the same level
as for the F(5008)/F(4960) ratio, the $\pm$ 3.3\% error would 
result in only a minor $T_e$ error, e.g., $\pm$ 100~K at $T_e$~= 10$^4$~K
(see \S5).
Because of the fringe pattern, we measured
the fluxes F(5008) and F(4960) over an integer number of fringing cycles
and then calculated the F(5008)/F(4960) ratio.
The overall value is 3.008.
When we apply a very small differential extinction correction (see \S4),
the intrinsic ratio becomes 3.000, which is in good agreement with the 
theoretical prediction.

\section{EXTRACTION of EMISSION LINE FLUXES from WFPC2 DATA}

	One of the most important operations involved
partitioning the contributions to the F437N emission. 
We consider these as due to [\ion{O}{3}] 4364, H$\gamma$, 
and continuum emission.
(An examination of our STIS spectra shows no other emission lines will
contribute significant flux.)
It is necessary to take into
account the total system throughput
for the F437N filter (Biretta et~al.\ 2001).
A finer scale digital version of the total system throughput for this or any
WFPC2 filter can also be produced using the task {\it calcband}.
We used a digital form to obtain the throughputs for both the 
[\ion{O}{3}] 4364 and H$\gamma$ lines accounting for the Doppler 
correction for the actual velocity of NGC~7009 at the time of observation.  
This was $-$20.71~km~s$^{-1}$ 
after accounting for the heliocentric velocity of 7009, which is 
$-$46.6~km~s$^{-1}$ (Schneider 1983).  
While still a net blue shift ($-$0.3~\AA), this does result in a slightly higher
H$\gamma$ leakage due to near maximum red shift of Earth's motion
at the time of observation.
The relative transmission at the Doppler
corrected wavelengths for H$\gamma$ to 4364 was 0.0253.

	In order to assess the contribution of 4364 and H$\gamma$ leakage 
to our F437N data, we did a careful spatial registration/comparison  
of the highest signal-to-noise (S/N) portions of our STIS long-slit data
with corresponding WFPC2 filters. 
Two areas were chosen that we call E1 and WW that were regions of
higher S/N and avoided the CS as well as the fiducial bars.  
They span from 1.05$''$ to 7.60$''$ east and from 0.75$''$ to 8.10$''$ 
west of the CS, respectively, along our 0.5$''$-wide STIS slit.
The registration is very close with the E1 area from STIS 3.275 
and from WFPC2 3.2737 arcsec$^2$.  
For WW, areas are respectively, 3.675 and 
3.6705 arcsec$^2$.

	Initially, we also tried to determine the continuum
contribution to F437N from the spatial registration of
our STIS data with the corresponding WFPC2 filters. 
All our efforts resulted in continuum fluxes that were
generally too large for what
could possibly fit for the WFPC2 observations.  
While it is not clear why the flux levels underlying the
emission lines (not just 4364 but also H$\alpha$, H$\beta$, and 5008) 
were invariably too high, the integration times
with STIS were insufficient to provide adequate
signal-to-noise to measure these continuum fluxes.

	Because the dominant contribution to the nebular continuum for NGC~7009
is recombination processes, we estimate the continuum
contribution to each of the WFPC2 filters from recombination theory.
These were provided independently by two of us (JPH and GJF) from
theoretical continuum models, with similar results.
For these, we assumed $T_e$~= 10,000~K,
N(He)/N(H)~= 0.11, and 
ionic fractions: N(He$^+$)/N(He)~= 0.88 and N(He$^{++}$)/N(He)~= 0.12,
which are essentially the NGC~7009 values in 
Kingsburgh \& Barlow (1994).
The scaled values adopted for the continuum emission relative to the 
H$\beta$ line emission (all intrinsic)
are 6.91E-4, 6.53E-4, 6.45E-4, and 5.65E-4~\AA$^{-1}$
for F437N, F487N, F502N, and F656N, respectively.
To obtain the total continuum contribution in the respective filters, one
multiplies by the appropriate ``rectwidth" (RECTW) 
31.994,  33.8, 	35.562, and 29.89 \AA\ obtained using the task 
{\it bandpar}.\footnote{Note that RECTW is \underbar{not} the 
``effective width of the bandpass" tabulated in 
Table 6.1 and also shown as dashed rectangles
in the Passband Plots of Filter Throughput 
(Appendix~1) of the WFPC2 Instrument Handbook (Biretta et~al.\ 2001).}

	This provides the predicted continuum correction for F487N of 
0.02207~I(H$\beta$), where I(H$\beta$) is the intrinsic flux.
Because both line and continuum are equally affected by extinction, we may 
also work with the observed fluxes without regard to extinction correction here.
To predict the continuum correction for F656N and F437N,
we scale the H$\beta$ result to the H$\alpha$ and H$\gamma$
lines using recombination theory (Storey \& Hummer 1995) 
assuming $T_e$~= 10,000~K and $N_e$~= 10,000~cm$^{-3}$, Case B:
I(H$\alpha$)/I(H$\beta$)~= 2.85
and I(H$\beta$)/I(H$\gamma$)~= 2.13.\footnote{For conditions
applicable to NGC~7009, we find using an online program 
(see Storey \& Hummer 1995) that the 
I(H$\alpha$)/I(H$\beta$) ratio used here will depart by less than 
1.75\% over the range  7500 $\le$ $T_e$ $\le$ 12,500~K and
10$^3$ $\le$ $N_e$ $\le$ 10$^5$~cm$^{-3}$.
The I(H$\beta$)/I(H$\gamma$) ratio will vary by $\le$1\%
over the range  7500 $\le$ $T_e$ $\le$ 15,000~K and
10$^2$ $\le$ $N_e$ $\le$ 10$^6$~cm$^{-3}$.}
This results in predicted continuum corrections for F656N of 
0.005926~I(H$\alpha$)
and for F437N of  0.04709~I(H$\gamma$).
The continuum relative to 4364 is obtained by multiplying the ratio
of STIS observed fluxes of H$\gamma$/4364 by the continuum/H$\gamma$ value 
of 0.04709.  
The differential extinction from 4364 to 4342~\AA\ is negligible for 7009.

	The continuum correction for F502N
relative to [\ion{O}{3}] 5008 is obtained by
multiplying the ratio of STIS observed fluxes of
H$\beta$/5008 by the continuum/H$\beta$ value of 0.02294.  
Because the 5008 line is the strongest of the set, the
continuum correction is extremely small; hence, it is
safe to neglect the very small differential extinction
from 4863 to 5008~\AA.

	While the above discussion is appropriate
for correcting for the continuum in the specific WFPC2 filters, if
comparison is made {\it between} individual filters, then an extinction
correction would need to be considered (see \S 4).

	Our original observing plan had been to use the WFPC2/F547M 
data to help determine the continuum emission.  
This does not appear to be a reliable method 
as judged by comparing our STIS and WFPC2 data.
For instance, let us consider area E1.
The observed F547M flux is 1.128$\times$10$^{-12}$ erg~cm$^{-2}$~s$^{-1}$.
Using the continuum predictions of the theoretical model mentioned earlier
for the mean wavelength 5468~\AA\ of this filter, the 
continuum emission at 5468~\AA\ relative to the H$\beta$ 
line emission is 6.25E-4~\AA$^{-1}$.
To obtain the total continuum contribution in this filter, we
multiply by the RECTW~= 638.58~\AA\  
and then multiply by the observed STIS H$\beta$ flux of
1.507$\times$10$^{-12}$ erg~cm$^{-2}$~s$^{-1}$
to arrive at 
6.014$\times$10$^{-13}$ erg~cm$^{-2}$~s$^{-1}$.
This is only 53\% of the observed F547M flux.
In the above assessment, we have neglected the small differential extinction.
Presumably the other $\sim$47\% is due to line emission.
We made a reality check of this number by using the digital 
version of the total system throughput for this filter
and the measured line fluxes in Hyung \& Aller (1995b). 
The full width at 0.01 of the peak transmission of the F547M filter extends from
5002--5997~\AA.
As above, we account for the actual Doppler velocity of the source. 
The result is that by far the bulk of the line emission in F547M is
due to the 5008 line;
the ratio of the total continuum flux (calculated as described) 
to the scaled, transmitted 5008 line flux is 2.43.
We also include the next most significant contributors:
\ion{He}{1} 5877, 
\ion{He}{2} 5413,
and
[\ion{Cl}{3}] 5519,39.
The estimated contribution then of 
lines is $\sim$35\% of the total F547M emission.
If the many other weaker lines were included, this value
would increase somewhat.
The main point of this exercise is to show that 
there can indeed be serious ``leakage" of 5008 line emission
into the F547M data for NGC~7009 and
no doubt for other PNs.

	In addition to accounting for the continuum in each
WFPC2 filter and the H$\gamma$ leakage in F437N, it is also
necessary to account for the contribution of [\ion{N}{2}] 
6585.23 and 6549.86~\AA\ emission into F656N.  
This is done by measuring the fluxes of these lines, as well
as that of H$\alpha$, in the areas E1 and WW.  
The method follows from that used for the H$\gamma$ leakage in F437N. 
The Doppler corrected wavelengths for the three lines
are determined and the relative throughputs 
applied for the F656N filter.  
These are 0.03645 and 0.4552 for the 6585 and 6550~\AA\ 
lines respectively relative to H$\alpha$.  
Note that although the 6585 flux will always be
larger than that of 6550, the latter contributes more
to the observed F656N flux.  

%% TABLE 1 ABOUT HERE

	Table~1 shows the details of the comparison with the STIS
overlay for E1 and WW for each of the WFPC2 narrow-band filters.
Column 2 has the ratios of the observed line flux with STIS to the 
corresponding narrow band filter flux with WFPC2.
Column 3 shows the flux ratio Contm./emission line, derived from recombination theory
and the WFPC2 filter profile. 
Columns 4 and 5 provide the observed line flux ratio derived from STIS data
as modified passing through the pertinent WFPC2 filter profile.
Column 6 then shows the Total (of the previous columns 2 up to 5)
for the STIS flux (including the theoretical continuum) divided by
the flux observed in the narrow-band filter.
The results for both areas are in close agreement; thus
we use the combined area E1~+ WW, which we call SUM, for our actual 
calibration to obtain observed line fluxes from the WFPC2 data.
In the case of F437N and F502N, the total of the STIS emission plus
inferred continuum exceeds the WFPC2 flux by a factor of 1.089 and 1.043
respectively, while for F487N and F656N, 
the total STIS emission plus continuum is less than the WFPC2 flux 
by factors of 
1.032 and 1.037 respectively.
  From the SUM entry line for F437N in Table~1, we see that the relative 
strengths of the components comprising F437N are roughly in the
proportions 0.71~:~0.19~:~0.10 for the 4364 line~:~continuum~:~H$\gamma$ 
leakage.

	We use the results for the SUM area to define 
our best relationships for determining the necessary 
emission line fluxes from the WFPC2 data.
Because the F437N corrections for the continuum and
H$\gamma$ leakage are tied to a measurement of a Balmer line flux,
we use H$\beta$ from the F487N image to measure its flux.
Thus we arrive at equations (1)~-- (4).

\begin{equation} 
F(H\alpha) = 0.9531~ {\rm F656N}~~,
\end{equation} 
\begin{equation} 
F(H\beta) = 0.9482~ {\rm F487N}~~,
\end{equation} 
\begin{equation} 
F(5008) = 1.041~ {\rm F502N}~~,
\end{equation} 
\begin{equation} 
F(4364) = 1.089~ {\rm F437N} - 0.03218~ 10^{-0.124~ c(H\beta)} {\rm F487N}.
\end{equation}

\noindent
Equation (4) requires further description.  
Both the H$\gamma$  leakage and continuum
corrections relative to H$\gamma$  are known for the specific
STIS-calibrated areas above.  
However, when we deal with other parts of the nebula observed with WFPC2 but
not by the STIS slit, there is no direct way of knowing the observed H$\gamma$  
flux.  
Therefore, we make these corrections in terms of the observed H$\beta$ flux,
which may be assessed via the F487N data.  
Again, we use the theoretical flux ratio H$\beta$/H$\gamma$~= 2.13 but must
correct for differential extinction between the two lines.  
In the presence of extinction, the observed H$\gamma$ line will be weaker 
relative to the observed H$\beta$ line (than the theoretical ratio).  
Therefore, the contribution to F437N will be less and thus the correction 
term should be less.  
The extinction is measured by the parameter 
$c$(H$\beta$), which is the 
logarithmic extinction correction at H$\beta$, discussed in the next section.
The constant 0.03218 in equation (4) was determined by requiring perfect 
agreement amongst all the pertinent observed fluxes for area SUM.

\section{EXTINCTION and REDDENING CORRECTION}

	Before deriving the $T_e$ distribution from either the
STIS or WFPC2 data, we first correct for extinction.
This is calculated by comparing the observed 
F(H$\alpha$)/F(H$\beta$) ratio with the theoretical ratio 
I(H$\alpha$)/I(H$\beta$).
Again, we use a value of 2.85 assuming $T_e$~= 10,000~K and $N_e$~= 
10,000~cm$^{-3}$, Case B (Storey \& Hummer 1995). 
The extinction correction is done in terms of
$c$(H$\beta$),
given by the relationship
\begin{equation} 
log [F(\lambda)/F(H\beta)] = log [I(\lambda)/I(H\beta)] - f(\lambda)~ c(H\beta),
\end{equation}
\noindent
where $f(\lambda$) is the extinction curve.
For the 4364, 4863, 4960, 5008, and 6565 \AA\ lines, the respective values for
$f(\lambda$) are 0.124, 0, $-$0.023,  $-$0.034, and $-$0.323 (Seaton 1979).
This leads to, 
\begin{equation} 
c(H\beta) = 3.096~ log [F(H\alpha)/F(H\beta)] -  1.408 .
\end{equation}

	The correction for extinction/reddening from observed
to intrinsic flux for the 4364 and 5008 lines is then given by,
\begin{equation} 
I(4364) = F(4364)~ 10^{1.124~ c(H\beta)}~~; 
~~~I(5008) = F(5008)~ 10^{0.966~ c(H\beta)}~~~~  .
\end{equation}
\noindent
Note that the first of these was used to derive equation (4).

	For the STIS data, we binned the pixels along the slit into 
tiles  that are 0.5$''$ square (matching the slit width).
The CS and STIS fiducial bars are excluded.
This produced $c$(H$\beta$) results for 45 tiles, which are the same set 
used later for the $T_e$ analysis.
The statistics without regard to any weighting for brightness are:
the average $c$(H$\beta$) is 0.1032; the median is 0.1038; and
the values range from 0.0239 to 0.1592.
The median and average are small and close to previous spectroscopic values,
e.g., 0.12 (Barker 1983 and references therein).

	For the WFPC2 data, we work with the individual pixels.
There were a few negative $c$(H$\beta$) values,  which
are replaced with a value of zero.
There also were spuriously high $c$(H$\beta$) values inferred in
the FLIERS (Fast Low Ionization Emission Regions), 
particularly on the west side.
The reason for this is that the 
[\ion{N}{2}] 6550, 6585~\AA\ lines are very much stronger in the FLIERS.
This can be seen from the paper 
by Balick et~al.\ (1998) in their colour image (fig.~1) and emission-line
profiles (fig.~3).
The leakage by the 6550, 6585~\AA\ lines into F656N enormously exceeds
the ``calibration" shown in Table~1.
The application of equation (1) overestimates F(H$\alpha$) in
the FLIERS, resulting in a spuriously high F(H$\alpha$)/F(H$\beta$) 
and thus $c$(H$\beta$), reaching as high as 0.456.
With the guidance of the STIS slit overlay, we chose to cap
the $c$(H$\beta$) value in the vicinity of the FLIERS at a value of 0.14.
This value is not crucial and could have been set between 0~-- 0.2
because extinction is low for NGC~7009.
As will be seen in the next section,
it matters little for the subsequent temperature analysis.

\section{ELECTRON TEMPERATURE DETERMINATION}

	The electron temperature $T_e$ is derived from the
intrinsic ratio I(5008)/I(4364)  using the following relation,
\begin{equation} 
T_e = 32,966/[ln (I(5008)/I(4364)) - 1.701]~~  .   
\end{equation}
\noindent
Effective collision strengths are from Burke, Lennon \& Seaton (1989)
for $T_e$~= 10$^4$~K.
Transition probabilities (A-values) are from Froese Fischer \& Saha 
(1985).
Note that this holds in the low-$N_e$ limit, which should be valid
for NGC~7009 where $N_e$ values are less than the critical
densities ($N_{crit}$) for these lines.
The lowest $N_{crit}$ $\sim$6.4$\times$10$^5$ cm$^{-3}$ for the 5008 line,
which is well above $N_e$ values determined (e.g., Hyung \& Aller 1995a,b).

%% FIGURE 1 ABOUT HERE

	For the STIS data, we continue with the analysis
using the 45 tiles along the slit described above.
Equations (6)--(8) are applied to the four emission lines to derive $T_e$.
Figure~1 shows the distribution of  $T_e$ vs.\ position along the slit,
This is shown relative to the closest approach of the CS to
our slit (recall the CS was placed off the slit edge).
The open circles represent the individual tiles plotted at their midpoint.
The dashed lines are a linear interpolation across the tiles
that were deemed to have unreliable measurements because of proximity
to the CS (tile \# 25--27) and the east fiducial bar (9--11).
There appears to be some symmetry with respect to the CS.
  From a more or less flat $T_e$~$\sim$ 10,800~K over the centre tiles,
there is a decrease to $\sim$~9500~K and then a rise again 
to roughly 10,800~K in both the east and west with increasing distance 
from the star.  The situation is less clear on the east side, partly
because of the lack of data at the east fiducial bar. 
In the next section, we will evaluate this  distribution
in terms of $T_e$ variations.

	A plot of  the observed relative [\ion{O}{3}] 5008~\AA\
flux is also shown in Figure~1.
This curve is displayed (unsmoothed) at the pixel level
providing 0.05$''$ spatial resolution along the slit.
There are some noteworthy oscillations between 
$\sim$8--13$''$ west of the CS.
There appears to be 4 cycles with a period between 
$\sim$1 -- 1.5$''$ per cycle.
These reach a factor of $\sim$2 brightness contrast
between adjacent local maxima and minima.
The structure is very repeatable when compared to the corresponding 
run of observed H$\alpha$ and H$\beta$ flux.
Because these oscillations do not 
appear to be correlated with the $T_e$ 
distribution and are not present in
the flux ratios of 5008 to the Balmer lines, they must be 
due to a column density (or $\int  N_e^2 dl$) modulation.
Such a pattern might result from episodic mass loss.  
We find very similar spatial behavior,
including this tantalizing oscillatory pattern,
from our WFPC2 F502N, F487N, and F656N data when
we overlay the STIS slit.
We note that this result from our 1-D cut is insufficient to
establish a case for episodic mass loss.

	Using the WFPC2 data, the path to $T_e$ is more 
uncertain than is the case with our STIS measurements.
However, the WFPC2 data cover a much larger area of the nebula.
Equations (1)--(4) and (6)--(8) are used to derive $T_e$.
In Figure~2, we present the $T_e$ map at a resolution of 0.1$''$.
The outer region with white and black speckles is omitted from
the analysis.
The S/N in F437N is generally too low out there to obtain reliable 
$T_e$ information. 
Nevertheless, with a hint of the ansae, this area helps visually identify 
the major axis. 
We note that by binning the data in the outlying areas, the S/N may be
improved.
For the purposes of this paper, we restrict the analyses to the interior
main body of the nebula with the highest S/N, where 
there is sufficient area to address temperature fluctuations in 
this nebula.

%% FIGURE 2 ABOUT HERE

	The high $T_e$ white region immediately surrounding the CS is the 
result of diffracted or scattered stellar continuum. 
Equation~(4), which corrects for the continuum in the F437N filter, 
does not account for stellar continuum emission.  
The result is a spuriously high F(4364) leading to the 
spuriously high values for $T_e$.
Further out, the temperatures show the red and yellow colour code.  
The pattern that looks like 
an ``H"-shape region,
or for the imaginative, the ``batman symbol",
appears to be point symmetric with respect to the CS.
This region is likely to be at the elevated $T_e$s indicated
because of the agreement in this central region with the STIS results shown in
Figure~1.
The full STIS long slit is shown superimposed on Figure~2.
The tiles along the slit are entirely within the inner higher S/N region.
The distances from the CS to the centres of the two symmetrical blue 
``lakes" that the STIS slit passes through are roughly 6.6$''$.  
That distance corresponds to $\sim$13 tiles on Figure~1, where 
the CS location is within tile 26.  Thus the troughs in the $T_e$ structure
in Figure~1 near tile 13 and 39 correspond with the blue lakes.
The temperatures also start to rise again further out 
as shown by the green colour.
However, there is not a prominent increase in $T_e$ with increasing 
distance from the CS to the levels indicated from the $T_e$ distribution
along the slit in Figure~1.

	The determination of $T_e$ from I(5008)/I(4364) is not sensitive to
the extinction correction for NGC~7009 because extinction is small.
We may demonstrate this with an example using a 
$c$(H$\beta$) of 0.1, which is close to the average value.
Let us also use a case that results in a representative temperature 10,000~K.
The observed 
F(5008)/F(4364) would be 153.543 to result in I(5008)/I(4364)~= 148.057,
which results in $T_e$~= 10,000~K by equation (8).
If $c$(H$\beta$) were respectively 0.0 or 0.2 instead with the same observed
flux ratio, the resulting 
$T_e$ would be 9891 and 10,112~K.
Thus were we to carry out this analysis with a constant
$c$(H$\beta$)~= 0.1, or no extinction correction at all,
the results will be hardly altered.

\section{FRACTIONAL MEAN-SQUARE TEMPERATURE VARIATIONS}

	Our STIS and WFPC2 analysis above 
present results in the plane of the sky.
The observations here do not address temperature fluctuation along the
line of sight, which may be characterized in terms of the average 
temperature $T_0$ and fractional mean-square $T_e$ variation ($t^2$) 
as defined by Peimbert (1967).  
\begin{equation} 
T_0~=~ {\int\ T_e\,N_e\,N_i\, dV
\over{\int\ N_e\,N_i\, dV}}~~,
\end{equation} 
\begin{equation} 
t^2~=~ {\int\ (T_e-T_0)^2\,N_e\,N_i\, dV
\over{T^2_0\int\ N_e\,N_i\, dV}}~~,
\end{equation} 
\noindent
where $N_i$ is the ion density $N(O^{++}$).
The integration in equations (9) and (10)
is over the column defined by each
pixel (or tile for the STIS analysis), and along the line-of-sight ($los$).
We are unable to measure the $t^2$ along the $los$ for any column
(cross section 1 pixel or tile).  
If there are $t^2$ along the $los$, we can say that
$T(4364/5008)$ $>$ $T_0$ (e.g., Peimbert 1967; Rubin et~al.\ 1998).

	For each pixel/tile, we have calculated $T(4364/5008)$.
Then the intrinsic flux I(5008), fully correcting F(5008) for extinction 
(see equ.\ 7), in each pixel/tile is used in conjunction with 
$T_e$~= $T(4364/5008)$ for that pixel/tile, 
and assumed constant along the $los$, to derive the following:
\begin{equation} 
I(5008) = K(5008) \int\ N_e\,N_i\,T_e^{-0.5}\,exp(-\chi/k\,T_e)\, dl~~
= K(5008, T_e) \int\ N_e\,N_i\, dl~~.
\end{equation} 
Here $\chi$ is the excitation energy above the ground state for the upper level 
of the 5008 transition, k is Boltzmann's constant, $K(5008)$ is known from
atomic data, and $K(5008, T_e)$ has finally incorporated the known $T_e$
factor with the atomic constants.
Here we again make the safe assumption of the low-N$_e$ limit 
(negligible collisional deexcitation) discussed earlier.

	Let us now define analogous terms to represent the average
$T_e$ ($T_{0,A}$) and fractional mean-square $T_e$ variation 
($t_A^2$) in the {\it plane of the sky}.

\begin{equation} 
T_{0,A}~=~ {\int \int\ T_e\,N_e\,N_i\, dl\, dA
\over{\int \int\ N_e\,N_i\, dl\, dA}}~~,
\end{equation} 

\begin{equation} 
t_A^2~=~ {\int \int\ (T_e-T_{0,A})^2\,N_e\,N_i\, dl\, dA
\over{T^2_{0,A}\int \int\ N_e\,N_i\, dl\, dA}}
= {\int \int\ T_e^2\,N_e\,N_i\, dl\, dA
\over{T^2_{0,A}\int \int\ N_e\,N_i\, dl\, dA}}\, -1~~,
\end{equation} 

\noindent
where $dA$ represents an element of surface area in the plane of the sky and 
the integration over $dl$ is for each pixel/tile along the $los$.

\subsection{WFPC2 Analysis}

	For this part, the image in Figure~2 was further rotated
to have the NGC~7009 major axis align with the x-axis.
We obtained from this an array of intrinsic fluxes I(4364) and I(5008) for
input to our own analysis code. 
  For each of the above set of pixel-by-pixel solutions, we have the
[\ion{O}{3}] $T_e$ (that assumed no $T_e$ variations along the $los$).
A circular region (radius of 10 pixels~= 1$''$) 
surrounding the CS is excluded from further calculations.
Additionally, we exclude pixels near the
periphery, where the S/N for the 4364 \AA\ line is poor.

%% FIGURE 3 ABOUT HERE

%% FIGURE 4 ABOUT HERE

	Calculations for $T_{0,A}$ and $t_A^2$
were performed using various distances 
from the CS  (R$_{outer}$). 
This was done several ways:
assuming concentric circular shapes and concentric
elliptical shapes, which is perhaps more realistic for this source.
The bottom-line results are virtually unchanged between these different
methods.
In Figures~3 \& 4, we present our results for
$T_{0,A}$ and $t_A^2$
for an elliptical shape with an axial ratio of minor/major~= 0.808,
matching our $T_e$ map (Figure~2).
The abscissa R$_{outer}$~= $\sqrt{ a_{outer}~ b_{outer} }$,
where 
$a_{outer}$ and $b_{outer}$ are the semi-major and semi-minor axes.
Out to the most distant elliptical contour used
in our analysis ($a_{outer}$~= 14.5$''$,  
$b_{outer}$~= 11.7$''$, R$_{outer}$~= 13.0$''$),
there are 52,995 good pixels with all [\ion{O}{3}] $T_e$ 
values between 5531~-- 16,382~K.

	The dashed lines pertain to values for $T_{0,A}$ and $t_A^2$
for the entire area interior to 
the ellipse with $a_{outer}$,  
$b_{outer}$ and is plotted at the single distance R$_{outer}$
(again excluding the CS) and represent cumulative results.
The solid lines pertain to 
$T_{0,A}$ and $t_A^2$ in 
the {\bf annular area between two adjacent ellipses separated by 5 pixels
(0.5 arcsec) along the major-axis}.
This is plotted as a single point at a distance 
R$_{centroid}$~= $\sqrt{ 0.5~ (a_{outer}~ b_{outer} + 
a_{inner}~ b_{inner}) }$~~.

There is nearly a monotonic decline in $T_{0,A}$ with increasing
distance (as characterized for the ellipses) from the CS.
The results indicate very small values~-- $<$ 0.012~-- for $t_A^2$ throughout.  
The overall $T_{0,A}$~= 9912~K, 
$t_A^2$~=  0.00360 for the 52,995 good pixels within 
the most distant elliptical contour used (excluding CS region).
There is a steep climb in $t_A^2$ in
the outer elliptical annuli
(solid line in Fig.~4), although this may be due to noisier data
at larger distances from the CS.

\subsection{STIS Analysis}

	Our STIS data allow an even more direct determination of $T_e$ and 
hence of $T_{0,A}$, $t_A^2$, albeit for a much smaller area than with WFPC2.
The average [\ion{O}{3}] temperature $T_{0,A}$ using 45 tiles 
(excluding the CS and fiducial bar) is 10,139~K; $t_A^2$ is 0.00350.
The median value (without regard to proper weighting) is 10,295~K.
All 4 emission lines peak in 
tile 37, with tile 38 a close runner-up.
This is where our slit crosses the ``ridge" near the blue--green 
(\ion{He}{2} 4687 -- [\ion{O}{3}] 5008)
boundary in the 3-colour image 
(Balick et~al.\ 1998) at roughly 5--6$''$ from the CS.
These tiles have relatively low 
[\ion{O}{3}] $T_e$ of 9823 and 9755~K respectively (see Figure~1).
No doubt, the enhancement in line emission here is due
to  a greater column  density (and/or higher density)
and not temperature.
As determined from the extinction-corrected
5008 surface brightness (equation 11),
these tiles have the highest ($\int N_e~N_i~dl$)--weighting in 
equations (12) and (13).

	We note that the single pixel (out of 52,995)
with the highest 5008 surface brightness 
also has the highest 
($\int N_e~N_i~dl$)--weighting,
($\int T_e~N_e~N_i~dl$)--weighting, and
($\int T_e^2~N_e~N_i~dl$)--weighting
in the WFPC2 analysis.
This pixel is 
located with respect to the CS
5.9$''$ along the west major axis and 4.7$''$ 
orthogonally below
and is associated with the ``ridge" near the blue--green 
boundary of the Balick et~al.\ (1998) picture.

\section{DISCUSSION AND CONCLUSIONS}

	We determined the electron temperature from the flux ratio of 
[\ion{O}{3}] (4364/5008) using both STIS and WFPC2 observations of NGC~7009.
The resulting $T_e$ values from both the WFPC2 map and the distribution along 
the STIS slit do not vary much, with almost all values between 9000~-- 11,000~K.
The higher $T_e$ values are closer to the central star and appear from
WFPC2 images to coincide with the inner He$^{++}$-zone.
The temperature is higher in the He$^{++}$-zone
because heating is stronger due to radiation beyond 54.4~eV, 
while the cooling is less efficient as O$^{++}$ gives way to O$^{+3}$. 
Models show that some O$^{++}$ 
remains inside the He$^{++}$ zone, so the
higher temperature there will be reflected in T(4364/5008)~-- 
see, for example, figs.\ 11 and 14 of Harrington et~al.\ (1982).

	The observations here do not address $T_e$ fluctuation along the line of sight.  
We assume for each square column (projection of 1 pixel for the 
WFPC2 data and 1 tile for the STIS data on the plane-of-the-sky) that the 
plasma along the line of sight is isothermal at the $T(4364/5008)$. 
The analysis of both data sets for the average $T_e$ and fractional mean-square 
$T_e$ variations in the plane-of-the-sky, which we call  $T_{0,A}$ and $t_A^2$, 
gave consistent results.
For the WFPC2 analysis, the preferred method used concentric elliptical 
contours to conform to the natural elliptical shape of the nebula.
For $T_{0,A}$, the STIS average was 10,139~K, while the WFPC2 overall average
to the outermost elliptical contour considered was 9912~K.
Interior to the contours closer to the central star, $T_{0,A}$ is higher 
as depicted in Figure~3.
In all the statistical analyses, a region around the CS is excluded.
For both STIS and WFPC2, $t_A^2$ is found to be very small,
$<$ 0.012 everywhere.
Using STIS, we obtained 0.0035, which is close to
the WFPC2 overall value of 0.0036 within the most distant 
elliptical contour used.

	Because the continuum correction for the F437N data 
is a major source of uncertainty in the process of deriving $T_e$ 
from the WFPC2 observations, we attempted to assess this using another 
approach.
Instead of estimating the continuum correction theoretically for F437N,
we assume that STIS and WFPC2 are perfectly calibrated.
By this we mean that the Total STIS emission/F437N in Table~1 should be unity
(instead of 1.089).
We then assume that the continuum emission 
contribution to the common WFPC2/STIS area SUM (see \S3 and Table~1) 
be the remaining amount after accounting for the STIS 4364 and H$\gamma$ emission
as they would be observed passing through the F437N/telescope system.
This reduces the Contm./4364 value for SUM in Table~1 to 0.1525
and alters the proportions that would comprise the F437N emission 
to be 0.77~:~0.12~:~0.11 for the 4364 line~:~continuum~:~H$\gamma$ leakage.
This treatment results in a modification of equation (4) with 
the factor 1.089 becoming 1.0 and the factor 0.03218 becoming 0.02316.

	We carried out the subsequent analysis with just this change to equation (4).
The results were that the highest $T_e$ values were lowered several hundred
degrees while the lowest $T_e$ values were raised several hundred degrees. 
Consequently, the effect was to decrease $t_A^2$ values somewhat.
Repeating the elliptical annuli analysis, we found 
the WFPC2 overall value of 0.00303 (decreased from 0.00360) 
within the most distant elliptical contour used.
For the individual annuli, $t_A^2$ $<$ 0.01 everywhere.
For $T_{0,A}$, the WFPC2 overall average
to the outermost elliptical contour considered increased slightly to
9940~K (from 9912~K).
Essentially, there were only minor changes to the curves depicted in
Figures~3 and 4.
We are led from the above computations
to conclude that our results are robust and not sensitive in
the situation considered here to the correction for continuum emission
in the F437N data.

	Fluctuations in $T_e$ along the $los$ are inevitable.
We can make some comments about how our results for
$t_A^2$ 
might be adjusted by $T_e$ variations along the $los$.
The relationship between $T(4364/5008)$ and $T_0$ is
\begin{equation} 
T(4364/5008) = T_0~[ 1 + 0.5 (91,200/T_0 - 3) t^2]~~,
\end{equation} 
(e.g., Peimbert 1967; Rubin 1969).
For a given column, let us take a representative\break
$T(4364/5008)$ of 10,100~K, which
is close to the average $T_{0,A}$ for the STIS analysis.
If $T_0$ were 9000~K, then $t^2$~= 0.0343 along that $los$ in the column;
if $T_0$ were 8000~K, then $t^2$~= 0.0625.
If we assume that generally a given $los$ column 
will be characterized by $T_0$~= 9000~K, 1100~K lower than
$T(4364/5008)$, then according to the left-side form of
equation (13), one would expect 
$t_A^2$ to be a factor of 1.26 larger. 
This is due to using the lower 9000~K temperature ($T_0$) in the
denominator instead of the 10,100~K temperature $T(4364/5008)$.
Our conclusion will still remain that $t_A^2$ is very small
as measured by the HST data here.

	We do not have the data here
to characterize $T_e$ variations in 3-dimensions (3-D). 
One way that this may be realized is to consider the following case.
If every single pixel/tile $los$-column were to have 
$T(4364/5008)$~= 10,100~K, $T_0$~= 9000~K, and $t^2$~= 0.0343,
the analysis for variations in the plane-of-the-sky that we present here 
will result in $T_{0,A}$~= 10,100~K and  $t_A^2$~= 0. 
However, the overall true 3-D $T_0$~= 9000~K and $t^2$~= 0.0343.
It is useful to define an {\it overall} 3-D average $T_e$ ($T_{0,V}$)
and fractional mean-square $T_e$ variation ($t_V^2$).
These single values apply for the {\it entire source}.
Equations (9) and (10) define these specific values when the integration
is over the entire volume.  
For the above simple, contrived case, then $T_{0,V}$~= 9000~K and 
$t_V^2$~= 0.0343.
We note that for a spatially unresolved object
(total integrated fluxes observed in the aperture), 
$t_V^2$~= $t^2$ and a calculation of $t_A^2$ is meaningless.

	Our measurements of $T_e$ reported here are an average
along each line of sight.
Because each element of area treated in the plane of the sky
represents a column which has already created a spatially averaged
temperature along the $los$
(e.g., see fig.\ 1 in Rubin 1969), it is likely that 
the value for $t_V^2$ is substantially higher than $t_A^2$.
Measurements of $t^2$ along various sight lines 
appear to be the most direct way to reliably gauge $t_V^2$.
Therefore, despite finding remarkably low $t_A^2$, we cannot completely
rule out temperature fluctuations along the $los$ as the cause 
of the large abundance discrepancy between heavy element abundances 
inferred from collisionally excited emission lines compared to those 
derived from recombination lines.
Further work, beyond the scope of this paper, is underway that
will use modeling as well as additional observational data
in an effort to better determine the relationship between
$t^2$, $t_A^2$, and $t_V^2$. 

Judging from both the shape of the inner parts of the nebula and 
echelle observations (Balick, Preston \& Icke 1987) of NGC~7009, the inner 
regions of the nebula are well modelled as the projected surface of a 
prolate ellipsoid.  If this is true, then most of the lines of 
sight of interest in this paper intersect two thin sheets of emitting 
gas.  Thus the temperature fluctuations along the $los$, if they exist, 
lie within these sheets.  The internal physical structure 
within each of these sheets is probably not resolvable either in 
extant images or through high-dispersion spectroscopic observations.

Hydrodynamic models suggest that 
this ellipsoidal sheet consists
of nebular material compressed by the supersonic 
growth of the hot (optically invisible) wind-heated bubble inside the 
surfaces.  Models predict that this compressed gas cools efficiently 
and, to first order, is isothermal except, perhaps, in a weak shock 
at the leading (outer) edge.  
Dopita (1997) has argued that the emission 
lines from such shocks are all but indiscernible since the integrated 
energy density of the shock is small compared to other energies that 
can excite the emission lines.  However, the trailing edge of the 
compressed sheet
is generally assumed to be a simple contact 
discontinuity (``CD'') across which no pressure gradient exists.  In 
these models, in which magnetic fields and thermal conduction are not 
considered, no $T_e$ gradient is expected across the CD.
Recent x-ray observations (Chu et al.\ 2001; 
Kastner et al.\ 2000, 2001) 
call this description of the CD into question.  In particular, the 
edges of the hot bubble in contact with the CD are much cooler and 
denser than expected.  It seems possible that heat from the hot 
bubble is being conducted from the bubble into the trailing edge of 
the sheet.
Hence 
from the bubble ($T_e \sim 10^6$~K) into the compressed gas ($T_e \sim 
10^4$~K) 
a steep temperature gradient
may exist
over a thin region.

\acknowledgments

We are grateful to Aaron Svoboda for creating the C$^{++}$ code
PIXHUNTER and to the 
NASA Undergraduate Research Program (NASA-USRP),
which made his 2001 summer work at NASA/Ames possible.
We thank the referee for useful comments.
Valuable contributions were made during early stages 
by Chris Ortiz and John Nguyen.
RHR acknowledges support from the Long-Term Space Astrophysics (LTSA)
program, NASA/Ames Research Center contract NCC2-9018 with Orion
Enterprises,
and thanks Scott McNealy for providing a Sun workstation.
Support for this publication provided by NASA through Proposal 
number GO-8114.
submitted to the Space Telescope Science Institute, which is operated by 
the Association of Universities for Research in Astronomy, Incorporated, 
under NASA contract NAS5-26555.

\vskip-0.2truein

\appendix

\vskip-0.2truein

\section{Description of PIXHUNTER}

This was a 2001 summer project of Aaron Svoboda working at NASA/Ames with RR.  
There are separate
procedures to identify and remove bad pixels from emission lines
and from the continuum.
For the continuum regions of interest, generally adjacent to emission
lines to be measured, sections of the rectified 2-D image are identified
that are reasonably flat in flux in both the spatial and dispersion
directions.  Sections are necessary in order to avoid the CS
and STIS fiducial bars.
PIXHUNTER works by computing   
the mean pixel value and standard deviation of a given section.    
After computing these values, it again goes through the image and outputs 
to a file the coordinates of any pixel whose flux value is further away 
from the mean than $\pm$ a specified multiple of the standard deviation
($\sigma$); here we used 5~$\sigma$.  
We linearly interpolate the good data to replace the bad pixel values.

	Though the above algorithm for finding the bad pixels 
works well for the continuum sections, it 
is incompatible with emission lines,  
because the flux in an emission line can vary 
substantially in both the wavelength and spatial dimensions.  
For cleaning an emission line, PIXHUNTER works interactively 
with {\it splot}.
We plot and compare a single column (fixed wavelength)
in the line at a time with one of the 
fitting functions available 
(e.g., a Legendre polynomial).
For each individual column in the line, this fitting function
represents the distribution of flux with spatial position along the slit.
Following the fit to the emission line distribution,
{\it splot} will flag all pixels that deviate by more than a specified multiple of 
$\sigma$ from the functional fit.  
Various keys/commands permit further processing such as 
interpolating all of the flagged pixels in a column to the function.  
When the resulting image is satisfactorily repaired, 
it can be saved back into the original 2-D rectified image. 
Each column can be treated in this fashion by scrolling 
through interactively using existing options/tools within {\it splot}.
There is another method for cleaning bad pixels in the vicinity of
pronounced peaks in the spatial distribution.  This involves taking the
ratio of a single column in the line to the {\it blkavg} of the entire line.
Again, {\it splot} is used interactively, operating on each such ratio,
in a similar fashion to the first method.

\clearpage




%	\bibitem[Auri\`ere(1982)]{aur82} Auri\`ere, M.  1982, \aap,
%	    109, 301

\clearpage

\large
\centerline{{\bf TABLE 1}}
\large
\bigskip
\centerline{{\bf Flux ``Calibration" of WFPC2 Filters with STIS Data}}
\medskip
\large
\begin{center}
\begin{tabular}{cccccc}
\hline
\hline

\multicolumn{6}{c}{F437N} \\

\hline

\multicolumn{1}{c}{Location}    &    \multicolumn{1}{c}{4364/F437N}  &
\multicolumn{1}{c}{Contm./4364}   & \multicolumn{1}{c}{4342/4364} & &
\multicolumn{1}{c}{Total/F437N} \\
E1  &    0.7834     &     0.2610      &     0.1402  & &    1.098\\
WW  &    0.7623     &     0.2728      &    0.1467   & &    1.082 \\
SUM &    0.7714     &     0.2677      &     0.1438  & &    1.089 \\
\hline
\hline

\multicolumn{6}{c}{F487N} \\
\hline
\multicolumn{1}{c}{Location}    &    \multicolumn{1}{c}{4863/F487N}  &
\multicolumn{1}{c}{Contm./4863} & & & \multicolumn{1}{c}{Total/F487N} \\
E1   &   0.9431     &     0.02207   & &   &   0.9639 \\
WW  &    0.9520     &     0.02207     & & &     0.9730  \\
SUM  &   0.9482     &     0.02207     & &  &   0.9691  \\
\hline
\hline

\multicolumn{6}{c}{F502N} \\
\hline
\multicolumn{1}{c}{Location} &        \multicolumn{1}{c}{5008/F502N}  &
\multicolumn{1}{c}{Contm./5008}  & & &  \multicolumn{1}{c}{Total/F502N} \\
E1   &   1.031      &     0.00201     & & &   1.033 \\
WW   &   1.048      &     0.001984    & & &   1.050 \\
SUM  &   1.041      &     0.001995   & &  &   1.043 \\
\hline
\hline

\multicolumn{6}{c}{F656N} \\
\hline
\multicolumn{1}{c}{Location}     &   \multicolumn{1}{c}{6565/F656N}  &
\multicolumn{1}{c}{Contm./6565}   &  \multicolumn{1}{c}{6550/6565} &
\multicolumn{1}{c}{6585/6565} &   \multicolumn{1}{c}{Total/F656N} \\
E1   &   0.9484     &     0.005926       &   0.00407      &  0.00106 &
0.9589   \\
WW   &   0.9564     &     0.005926        &   0.005538     &  0.001361 &
0.9687    \\
SUM  &   0.9531     &     0.005926      &   0.004913     &  0.001231 &
0.9646   \\
\hline
\hline

\end{tabular}
\end{center}

%% FIGURE 1 
\begin{figure}
\plotone{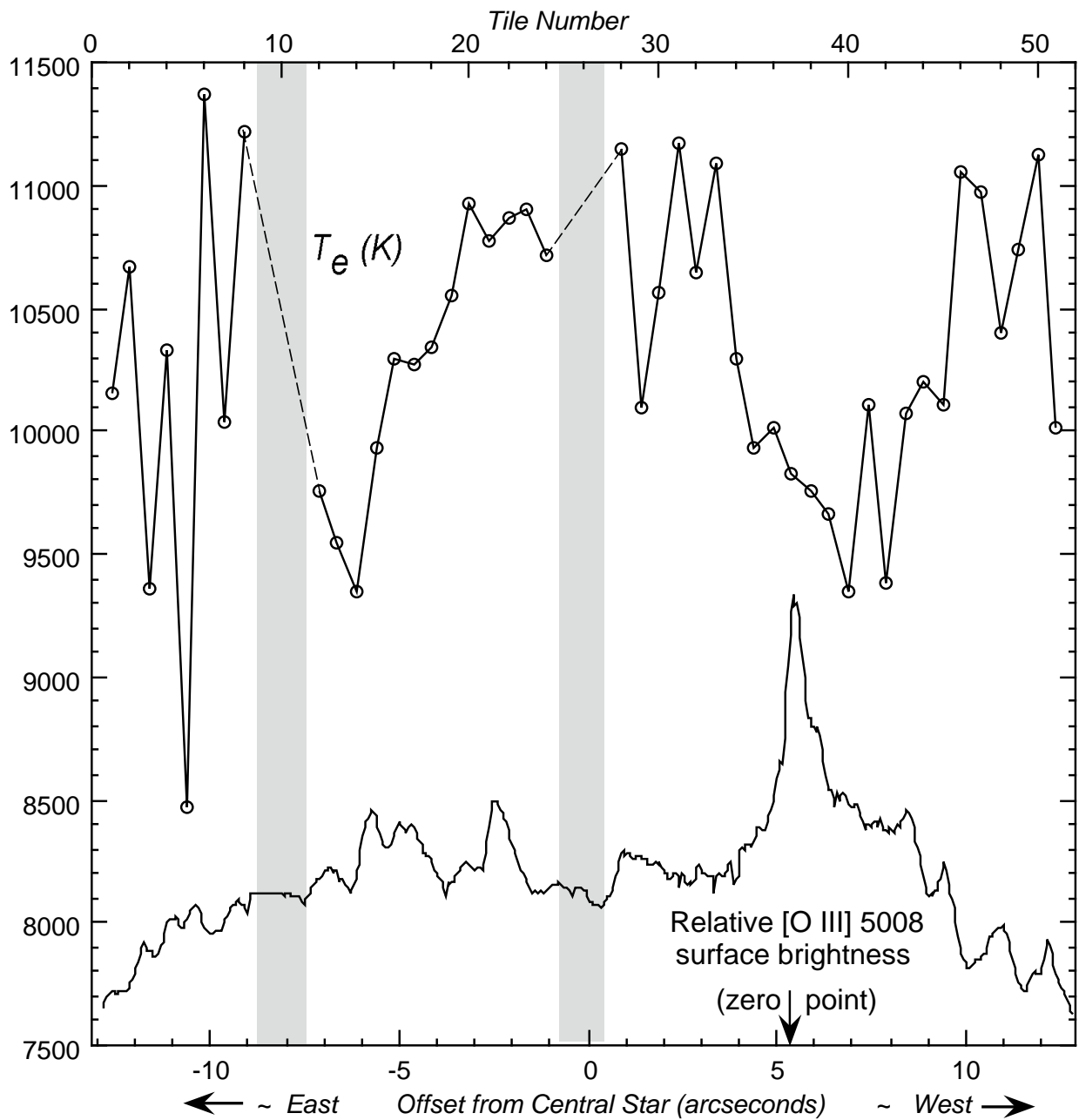}
\caption{Plot of $T_e$ determined from the [\ion{O}{3}] 4364/5008 flux ratio
vs.\ position along the STIS long-slit.  The analysis is in terms of
tiles  that are 0.5$''$ square (matching the slit width).  
There are  45 tiles with reliable measurements that exclude the gray areas.
The open circles represent the individual tiles plotted at their midpoint.
The dashed straight lines are
interpolations across the central star tiles (\# 25--27) and across
the east fiducial bar tiles (9--11).
Tile \# increases from $\sim$E to $\sim$W.  
The bottom curve shows the observed relative [\ion{O}{3}] 5008~\AA\
flux displayed unsmoothed at the pixel level.}
\end{figure}

\clearpage

%% FIGURE 2 
\begin{figure}
\plotone{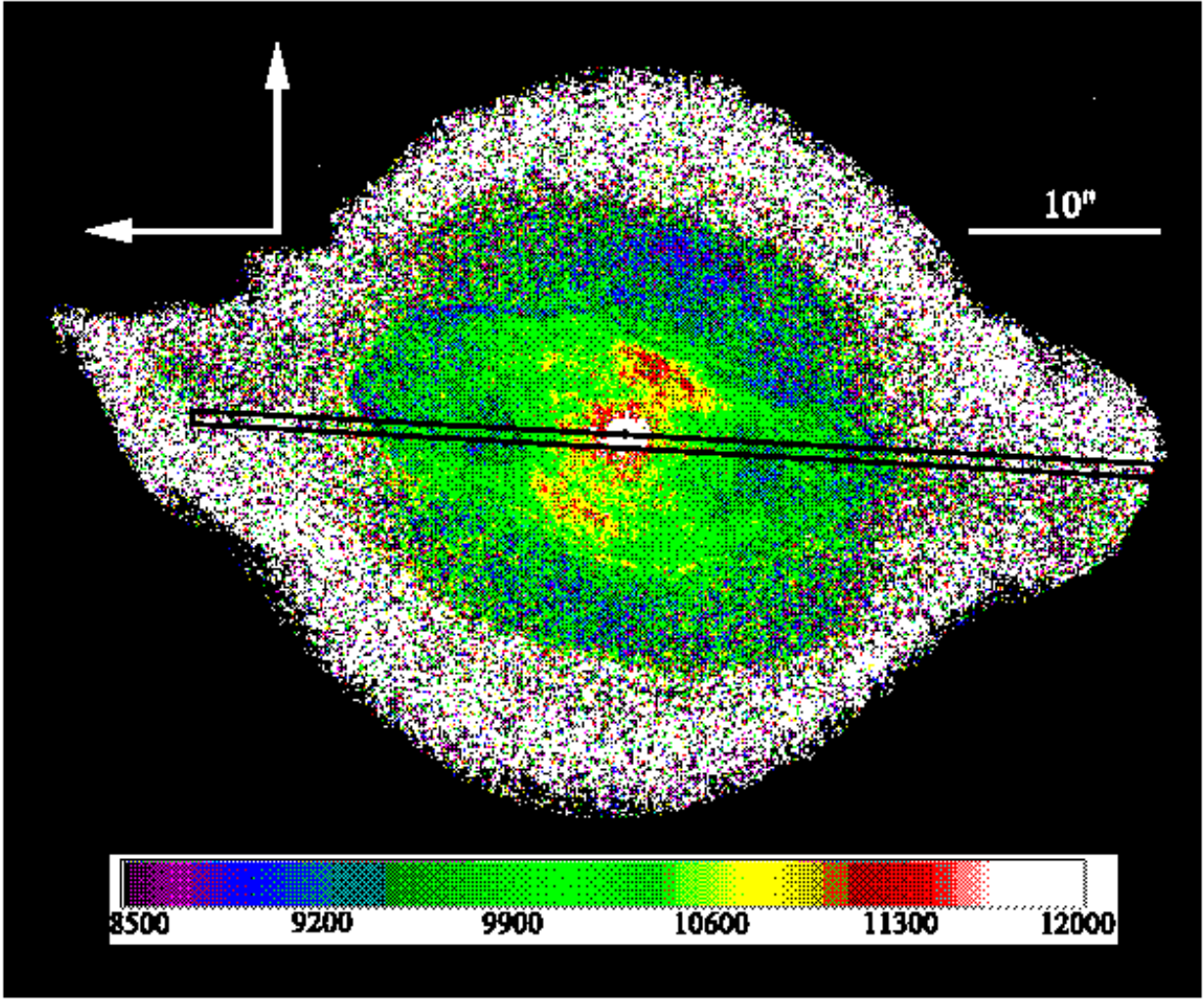}
\caption{This is a pixel-by-pixel map of the electron temperature $T_e$
(colour-bar scale) as determined from the ratio of extinction
corrected fluxes of  [\ion{O}{3}] (4364~\AA )/(5008~\AA ).
The STIS long-slit (52$''$ $\times$ 0.5$''$) 
is aligned and superimposed on the image.  
The area representing the 0.5$''$-wide slit is shown fully unobstructed.
The circumscribed black boundary lines are not part of the STIS slit.
N is up and E to the left.}
\end{figure}

\clearpage

%% FIGURE 3 
\begin{figure}
\vskip-2.0truein
\plotone{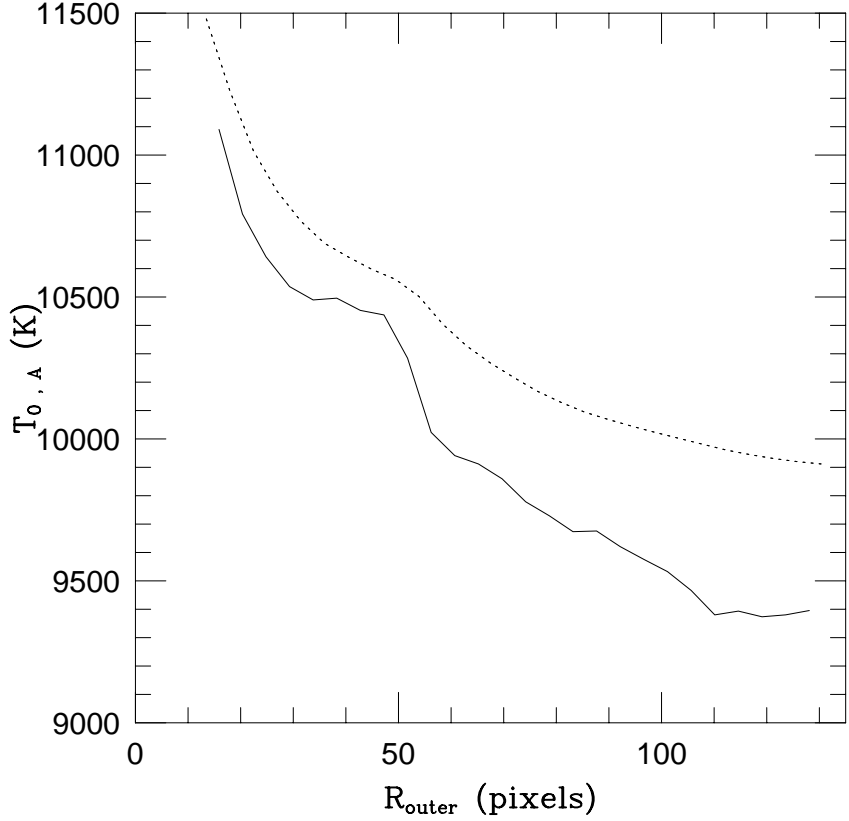}
\vskip-0.5truein
\caption{$T_{0,A}$ vs.\ R$_{outer}$~= 
$\sqrt{ a_{outer}~ b_{outer} }$.
R$_{outer}$ is measured from the central star
in pixels with 10 pixels equal 1 arcsec.
The dashed curve pertains to values for $T_{0,A}$ 
for the entire area interior to the ellipse with $a_{outer}$,  
$b_{outer}$ and is shown as a point at R$_{outer}$. 
The solid line pertains to $T_{0,A}$ in 
the annular area between two adjacent ellipses.
This point is plotted at the annulus area centroid (see text).
The step along the major axis 
between two successive elliptical
annuli is 5 pixels; along the minor axis, the step is 4.04 pixels.}
\end{figure}

%% FIGURE 4 
\begin{figure}
\vskip-2.0truein
\plotone{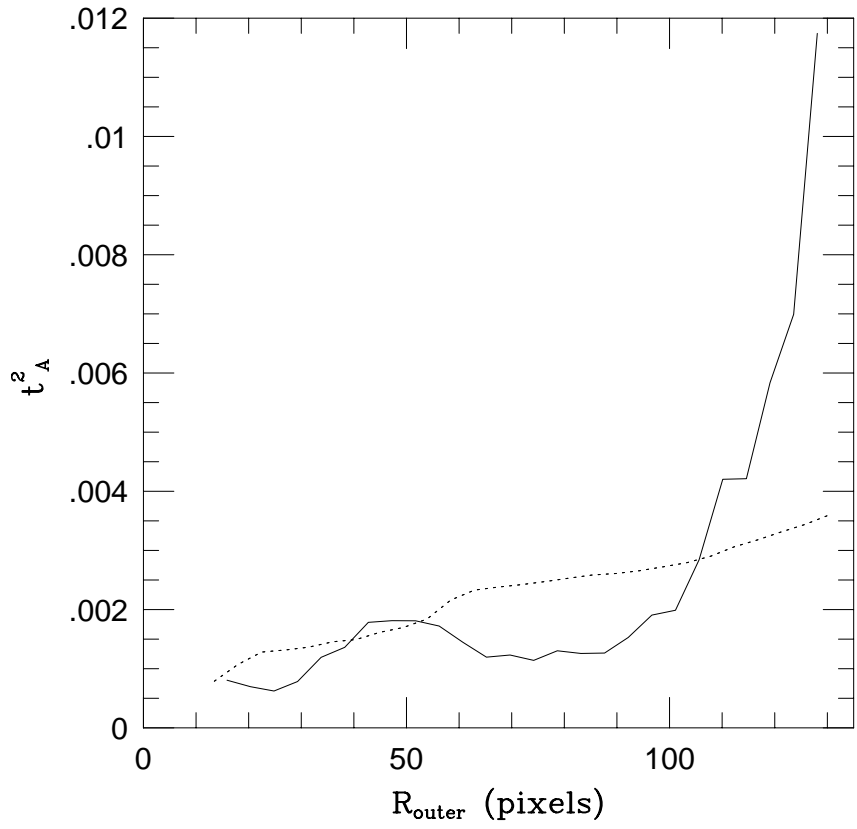}
\vskip-0.5truein
\caption{Same as Figure~3 except for $t_A^2$
instead of $T_{0,A}$.}
\end{figure}

\end{document}